# Mobile Web – Strategy for Enterprise Success

Jitendra Maan

Tata Consultancy Services, TCS Towers, 249 D & E, Udyog Vihar Phase IV, Gurgaon, Haryana, India – 122001

`Jitendra.maan@tcs.com`

*Abstract.*

*Today, enterprises are faced with increased global competition in an environment where customers are demanding faster delivery, better service and also want to gain significant and immediate business value by increasing productivity and reducing operational cost.*

*Spurred by unprecedented customer demand, each Industry cluster has developed its own source of comparative advantage. Even within a single organization, the business value chain is geographically fragmented. Such diversification and fragmentation of value chain drives the need for cross-platform Web applications over mobile channel. Mobile Web is the next logical transition in this evolutionary process and Mobile Web applications will continue to gain more prominence in the enterprises not just to improve the return on investment in their existing system landscape, but also to expand global reach and improve operational efficiency of their mobile workforce.*

*This paper outlines the critical business needs to rapidly create flexible Mobile web solutions across all lines of business. The paper enlightens the benefits offered by enabling web applications on Mobile devices and also addresses the current business challenges in developing Mobile Web applications.*
*This paper is intended for all business domains irrespective of application portfolios.*

## 1   Introduction

Due to proliferation of smartphone and featured mobile devices, the mobile technology landscape is evolving rapidly. Innovation in mobile technology, platforms and devices continue to grow and is bound to drive demand for extending enterprise applications and web content to the mobile devices. Mobile Web, not only brings a paradigm shift in the way the business applications are developed, delivered and consumed, but also changing the way, the businesses serve their customer. In such a competitive economic environment, it has become a business imperative to extend/enable enterprise applications and web resources to mobile devices.

However, in the past, technology was playing a crucial role in deciding on, how and where to access information, but today, customer want to gain the operational and cost advantages of deploying rich mobile applications over the Internet.

Today, enterprises need to align their technology practices and to instill the right composition of mobile technologies, platforms and disciplines in order to consistently execute ahead of their competitors. With such goal in mind, Mobile Web Solution acts as a key enabler to help customers in reducing operational costs, improve customer service and achieve a new velocity by providing real time visibility & monitoring into their operational processes with better data granularity.

It is also observed that Mobile Web access continue to grow with increased penetration of emerging mobile platforms, devices (like larger form factor Smart-phones, tablets) and continuous richness of micro browsers. Developing and deploying Web applications for different mobile devices is not as straightforward as it might sound. Due to proliferation of mobile devices, Mobile Web applications would have to be supported on a most of these devices running with different browsers. There is little or no coherence between different browsers supported by different Mobile platforms.

## 2  Mobile Web Paradigm – Challenges and Constraints

There is an ongoing challenge in front of CTOs/CIOs/IT Heads to develop an understanding what to mobilize, when to mobilize, as well as how to develop and execute mobility strategy in the context of their overall business eco-system. As business users become more mobile, enterprises face challenges to provide mobile applications to their workforce at the point of activity.

### 2.1  Key Business Challenges

Mobile Web paradigm is different from the normal client-server based application development. Therefore, the technical issues and design considerations are quite different in nature. The bottom line is that mobile Web application development is new to most organizations and comes with unique challenges and several un-answered questions. Some of them are listed below –

- Ongoing operational costs eating up a major part of customer IT budgets
- Lack of Mobile strategy to expand global reach & improve operational efficiency of mobile workforce
- Real-time access to customer and critical enterprise information is not available
- Limited standardized Mobile Web solutions available in the market
- Mobile Web Fragmentation

    – Mobile device diversity & platform fragmentation
    – Thousands of Connected devices
    – Multiple Operating Systems
    – Multiple Browsers
    – Multiple Form factors
    – Multiple Input methods

- How to extend and grow mobility solutions as the business needs change?
- How to utilize/re-use existing web oriented infrastructure to capitalize on Mobility opportunities?
- How can mobile Web improve customer satisfaction?
- What services and field operations require greater visibility?
- How to stay competitive and agile in changing business eco-system where Mobile domain plays a very crucial role.

## 2.2 Mobile Web – Technical Constraints

Mobile Web Applications, in general pose certain unique requirements and challenges, compared to their desktop versions, which primarily arise from -

- **Screen real estate** - Small Screen size with low/limited resolution
- Different browsers render applications differently on different devices due to varying capabilities
- **Hardware constraints** – Limited processor capabilities, limited RAM, Storage, physical size
- **Data usage** - High cost & slower data transfer rates,
- Limited battery Life is another barrier
- **Variable Network Connectivity** - Limited bandwidth, network availability
- **Interoperability issues** - Platform OS and device fragmentation
- **Usability issues** – Device Form factor (, user operating limitations
- **Limited inputs/interactions capabilities-** QWERTY keyboard , Touch Screen, Virtual Keypad
- Mobile browsers have varying capabilities like some may support handheld CSS, some may only support WML and some rich devices comes with HTML 5 support
- Most of the mobile devices support only a limited set of image and multimedia formats—e.g. Animated GIFs are not supported on most phones.

## 2.3 Mobile Web – Design Considerations

Mobile Web adoption would help to improve overall business user experience in the overall value chain. Here are the few factors that would be considered while designing Web application for mobile devices –

- **Design from the ground up** to take advantage of the new opportunities that mobile offers
- **Understand context of Mobile users -** One of the essential things while developing/designing for the Mobile Web is to understand the context of the user. Mobile context is different and is fundamental to the mobile UI strategy.
- **Understand changing user behavior –** Define viewpoint for the Mobile user
- **Simplified User Interface** - Keep the user interface as simple and intuitive, as possible.
- **Content -** Issues like navigation, image sizes, page weight and scripts all need to be considered when thinking about web application on mobile devices.
- **Full Web Browsing –** Delivering a full desktop like browsing experience to the user
- **Authentication and Authorization strategy** - Every design decision must take into account an effective authentication and authorization strategy for all possible mobility scenarios – Fully connected and occasionally connected modes.
- **Caching to improve response time** - Mobile Application design should factor-in caching as a mechanism to improve performance and responsiveness of mobile application. There is a need to choose right caching strategy including cache expiration and cache flushing policies.
- **Always design for mobile first** – Don't just re-purpose your web application
- **Adaptive rendering –** Browsers supporting most of the mark-up languages and mobile devices
- **Interactive and fluid UI interface –** Reduced page reloads using AJAX technology. Mobile user can retrieve only the data that he wants from the server, and change only the part of the page that he wants to change.
- **Content adaptation or multi-serving Strategies -**Content adaptation strategies would help to deliver content as per user device capabilities. For example, if the device is an old phone supporting only WML, then WML page is delivered to the mobile user along with Wireless Bitmap (wbmp) images and if it is a newer XHTML MP-compliant device, then an XHTML MP version is delivered & customized according to the screen size of the device.
- **Progressive enhancement philosophy** – Mobile browser uses all those layers (like HTML, CSS and JavaScript layers) which it can handle easily. A few decisions play a vital role to decide on which feature to support and/or leave out for which browser.
- **Support for Open APIs -** Open APIs are being exposed, ideally suited for mobile web and cloud apps, but they require appropriate monitoring and management
- **Over-The-Air (OTA) deploy-ability -** Mobility environment is highly heterogeneous, Over-The-Air provisioning/delivery of applications plays an important role as it enables easy deployment and upgrades to mobile web applications.

Majority of Mobile Applications contain some level of common functionality that
spans across layers and tiers. It is customary to examine the mobility functions required in each layer, and then abstract the functionality into common components to address cross-cutting concerns. And, such common components can be configured depending on the specific requirements of each layer of the mobile application.

The figure below shows different Mobile Application Models – Native Mobile
Applications, Mobile Web Application and Hybrid Mobile Applications. Such Model helps architects and developers to take right design decisions for cross platform strategy.

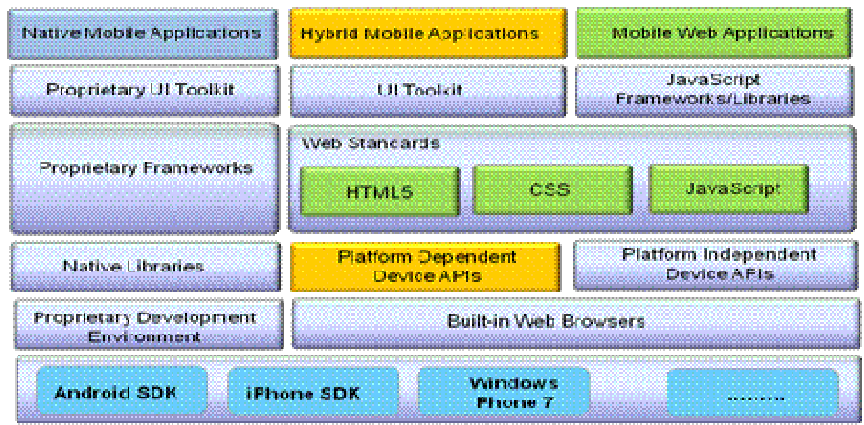

Figure1: Mobile Application Models

## 3   Mobile Web – Key Value Proposition

Using mobile web technology, companies are getting more out of their existing web-oriented systems by delivering applications and web content directly to mobile users. Moreover, it allow customers to achieve economies of scale in a global operation whereas, speed advantage results from enabling better and faster management decisions in business processes, which, in turn, result into easy and quick access to resources, real-time collaboration  and timely access to the right information.

### 3.1   Business Benefits

The key business benefits of Mobile Web Solutions are given below:-

International Journal on Web Service Computing (IJWSC), Vol.3, No.1, March 2012

- **Cross-platform Support -** Cross platform mobile Web solution would bring enormous benefits to business users to write-once-run-anywhere applications for a wide variety of mobile devices. By using HTML tools and a mobile application development platform, developers can write once and deploy to multiple platforms instantly.

- **Low upfront development cost -** Thin client approach and Web standards help to achieve high ROI from the mobile application, as there is only server side component development and testing of application on different browsers. In case of thick clients, there exists a need to have application development efforts both at the server side and client side; thus increasing the overall development and deployment cost.

- **Improve customer service and operational efficiency –** Mobile Web solutions address the typical business use-case to improve operational efficiency of mobile workforce by automating job scheduling, assignment, approval workflow and intelligent reminder/alerts triggered on cross platform devices.

- **Easy maintenance and Low TCO through Standards-based development** - Applications developed using Web standards are easy to maintain and update, resulting in low TCO for mobile Web applications in a long term. Offering would help developers to leverage existing Mobile Web standard tools that would rapidly integrate mobile applications with back-end and legacy applications and maximize the usability and flexibility of mobile applications.

- **Always Available Mobile Applications -** Always available mobile Web applications generate a higher rate of return due to the fact that users spend more time being productive with the application. Such Solution would be using **"always available architecture"** which clearly meant that users can access databases, applications and Web content quickly and easily on their mobile devices all the time.

- **More sales lead through interactive Mobile Web applications** - Interactive mobile web app development would make potential customers, interested and connected, leading to increased sales.  Collaboration tools would give them the feeling that they are a part of business eco-system.

- **Mobile Web offers a seamless user experience** - Using smart caching and advanced synchronization techniques, mobile Web Solution would deliver a seamless user experience across a variety of public and private wired and wireless networks

## 4  Mobile Web – Key recommendations

The Key recommendations, that may be considered while developing the Mobile Web applications are given below –

- Understand the customer  mobile platform preferences
- Select the platforms that are used by key target segments
- Have a clear business case for multi-platform mobile development
- Take into account the unique nature of mobile application development
- Keep as much of the logic in the network-side as possible
- Mobility solutions must utilize the existing Web-oriented infrastructure by reusing Web Services architecture as much as possible.
- Test the mobile web application on actual devices, with real customers, using it in the typical and expected real-world contexts/situations.
- Develop rich mobile web applications while minimizing costs using -

- Right architectural approach
- Deep knowledge of mobile as a medium
- Competitive delivery models

## 5   Benefits of Mobile Cloud Computing

Cloud computing is mainly the internet-based computing, whereby shared resources, software, and information are provided to computers and other mobile devices on demand, same as with the electricity grid. Mobile-enabled applications keep their data in the cloud and keep it updated with all the changes made.

Mobile Cloud Computing delivers significant value across the entire mobile ecosystem. Mobile Cloud Computing offers the organizations, an easy way to access capabilities such as Direct2Mobile Billing, click-to-call/conference, and 2-way messaging, seamlessly across multiple network operators. There are several benefits of deploying cloud services over mobile channel -

- **Cloud becomes a new medium to share resources and applications through Mobile channel** - As users become more mobile, the applications they use will become mobile-enabled. By hosting data and applications in the cloud, mobile users do not need to worry about to actually carry and maintain the data but such cloud driven model allow them to access their data & applications in real-time, anywhere on their mobile devices.

- **Create New Stream of revenues** - Several analysts predict that enterprise-scale cloud services based model will propel the growth of traditional IT service market using Mobile Channel. Due to nature of cloud applications, Mobile Cloud computing has started creating a new stream of revenue for businesses and application vendors since business adoption of cloud based services and platforms is growing at phenomenal pace.

- **Foster effective collaboration among applications and people**. The power of cloud allows deploying different collaboration tools (e.g. email, web conferencing, instant messaging and team workspaces etc) at centralized, high speed distributed servers to foster effective collaboration among different stakeholders. Take, for example, email services which has been cloud-enabled, will allow user to update mails on one device, remove a few emails using another device, and update calendar on $3^{rd}$ device and all of these devices are in complete sync using cloud services which offers a unique and connected user experience with up-to-date information in real-time. Solutions such as Microsoft's SharePoint, IBM's Lotus Live suite, and more recently Chatter from Salesforce.com, all aim to make it easier for employees to collaborate, whether they are producing a document, managing project tasks or simply wishing to communicate in real time as a group. Moreover, these solutions are increasingly being delivered as a cloud-based service and accessed via a desktop browser or mobile application.

- **Automatic backup of personal and enterprise data**. With cloud computing, you never lose your enterprise as well as personal data, because it always exists in the cloud. Since data would always be pulled from cloud stores, hence it becomes easy to replace one device with another, by simply registering a new device and get all your data from the cloud with no delay.

- **Easier and unique way to access computing resources.** Cloud makes it easy to login to various computing devices (e.g. desktop computer, notebook, tablet, cell phone, eBook reader) through single-sign on, a unified gateway to access all computing resources stored in the cloud environment.

- **More computing power on demand**. Cloud Computing offer the choice to select right services and deployment models in the interest of mobile users. Cloud computing opens up new innovative ways to use rich-media capabilities, such as integrating video into documents or presentations and computing power will become more and more disembodied and will be consumed as and when it is needed.

- **Offer bundled and unbundled services -** Cloud computing opens up new innovative ways to offer services bundled with video streaming and rich-media capabilities, such as integrating video into documents or presentations. And, the selection of such bundled or unbundled services over mobile domain, depends on various factors such as:

    1. Differentiated vs Commodity capabilities
    2. Core vs. non-core services
    3. Desire for flexibility and innovation vs. agility

We, at TCS believes in delivering on-demand business capability through "IT-as-a-Service" model, with an integrated suite of hardware, network, software and mobile solutions as a bundled service offering. More importantly, all the required technical, business and consulting services are provided in a "build-as-you-grow" or "pay-as-you-use" model through a combination of on-premise and shared services hosted platforms.

## 6     Challenges with Mobile Cloud Computing

### 6.1    Mobile Cloud inter-operability and Usability

Successful development and deployment of mobile Web applications demands for a better understanding of mobile user context and their challenges. With the screen resolutions, browser capabilities and speed (3G, 4G etc) are getting better now; mobile web experience is getting closer to desktop browsing. Generally, interoperability issues would stem from the platform fragmentation of mobile devices, mobile operating systems, and browsers.

From an engineering view of point, interoperability problems have to be addressed up-front for accessing cloud capabilities over mobile devices. Such problems would normally arise due to different form factors across mobile devices. Usability problems would be centered around the small physical size of the mobile phone form factors (limited resolution screens and user input/operating limitations.

Open Standards based interoperable services are necessary to preserve enterprise user data. The issue of mobile user freedom is surfacing with the emergence of more & more Mobile application stores available in the Cloud. Mobile Cloud success depends on interoperability among different computing devices like - mobile devices, desktop PCs and different app stores. For example – Syncing pictures with desktop PC and streaming videos from another Store; thereby giving unprecedented freedom to mobile users.

### 6.2    Network dependent factors drive mobile user experience in the cloud

Mobile cloud computing performance is dependent on various network factors like:-

- Network Latency
- Data transmission rate

It is customary to offload computer-intensive resources in the 'cloud,' and it requires special considerations in network design and application deployment.

### 6.3    Limited battery life of mobile device is another barrier

Another significant barrier in mobile cloud computing is the limited battery life of mobile devices. The more we execution in the cloud and more we save battery life, as the application execution burden is offloaded. Such execution offload is linked to device functions and cannot be completely transferred to the cloud.

### 6.4    Safeguard critical enterprise data in the cloud becomes an issue

It is not an easy task to safeguard enterprise user data sitting in the public/private cloud. It is necessary to develop Cloud based framework for effective collaboration and information management, sharing and archiving by taking into account the mobile user needs.

## 7    Summary

It is quite obvious that with the emergence of Rich Mobile Internet applications and Cloud technologies, customer's immediate focus has shifted towards those tools, technologies or platforms that deliver rich user experience that is visibly different from what's delivered by traditional server-centric platforms.

Clearly, Cloud as a mobile delivery model is fast shedding its nascent image with many organizations examining the possibility of employing Mobile Cloud for operational agility. Adoption of a cost-effective Cloud model is on the rise across businesses. Using such model, it is extremely easy to create an entirely new breed of products and services on mobile channel that deliver previously unheard of value to customers and this capability has been one of the significant pointers to the relevance, effectiveness, and sustainability of Cloud services.

Though, there are challenges in deploying Applications and data on the cloud, but the benefits that businesses can accrue in long term, simply cannot be ignored. There are obvious reasons to complement the Mobile enablement.

I strongly believe that Cloud-based mobility solutions will come to be recognized as an increasingly viable option for enterprise IT. In the current market scenario, even though the benefits of early Cloud adoption are well understood, there still exist numerous reservations pertaining to Cloud's capacity to support business needs, applications and underlying infrastructure. Early adopters with the insight to address such reservations will reap Cloud benefits in the long term..

## 8    Conclusion

Mobile Web has becoming a disruptive force in the mobility solution space. The strengths of the Web development paradigm are perfectly suited to creating mobile enterprise applications with a low total cost of ownership and a high return on investment. Mobile web not only offers unique and adaptive user experience using internet connectivity based models but also opens up social communication channels in the organization.

Though, there are challenges in Mobile Web application adoption, but the benefits that businesses can accrue in long term, simply cannot be ignored. It is obvious that with the emergence of Rich

Mobile Web applications and technologies, customer immediate focus has shifted towards those tools, technologies or platforms that deliver rich user experience that is visibly different than what's delivered by traditional server-centric platforms. In a nutshell, Mobile Web is emerging as a low-cost solution with compelling total cost of ownership (TCO) advantage over thick client applications.

## References.

[1] John Arne Saeteras, Mobile Web vs. Native Apps, Revisited, 2010
[2] Dana Moore, Raymond Budd and Edward Benson, Professional Rich Internet Applications: AJAX and Beyond, Wrox Press © 2007 Citation
[3] Designing for the Mobile Web, Sitepoint -http://www.sitepoint.com/designing-for-mobile-web/
[4] Forrester Report - Making The Case For The Mobile Internet by Julie A. Ask and Seth Fowler with J.P. Gownder and Laura Wiramihardja
[5] Gartner Report - Key Issues for Mobile Applications, 2010, William Clark, Publication Date: 25 June 2010, ID Number: G00201576
[6] Gens, F.: IDC Predictions 2010: Recovery and Transformation, Filing Information: IDC #220987, 1, 3-7 (2009)
[7] Gartner Report - Key Issues for Mobile Applications, 2010, William Clark, Publication Date: 25 June 2010, ID Number: G00201576
[8] Sudesh Prasad "Emerging Trends Enterprise Mobility: Always Connected", Voice & Data, First Quarter, 2006
[9] Kyung Mun, Corporate Technology Strategist, Alcatel-Lucent: Mobile Cloud Computing Challenges, September 201, TechZine, Technology and Research e-zone (September 2001), http://www2.alcatel-lucent.com/blogs/techzine/2010/mobile-cloud-computing-challenges